\documentclass[graybox]{svmult}

\usepackage{mathptmx}       
\usepackage{helvet}             
\usepackage{courier}            
\usepackage{type1cm}         
\usepackage{makeidx}         
\usepackage{graphicx}        
\usepackage{multicol}        
\usepackage[bottom]{footmisc}
\usepackage{amsmath}
\usepackage{natbib}
\usepackage{amsbsy}
\usepackage{color} 
\usepackage{hyperref,url}
\usepackage{ctable}
\newcommand{\teff}{\ensuremath{T_{\mathrm{eff}}}}
\newcommand{\logg}{\ensuremath{\log g}}
\newcommand{\feh}{\ensuremath{[\rm Fe/\rm H]}}

\newcommand{\nn}{n_{\rm non-LTE}}
\newcommand{\nl}{n_{\rm LTE}}
\newcommand{\mfp}{l_\lambda}
\newcommand{\bi}{b_i}

\newcommand\wlrange[2]{$#1 \le \lambda \le #2$}
\newcommand\ion[2]{\textup{#1\,\textsc{\lowercase{#2}}}}

\makeindex     

\begin{document}

\title*{Non-LTE radiative transfer in cool stars.}
\subtitle{Theory and applications to the abundance analysis for 24 chemical
elements}

\titlerunning{Non-LTE radiative transfer in cool stars: theory and applications}
%
%
\author{Maria Bergemann and Thomas Nordlander}
%
%
\institute{M. Bergemann \at Institute  of Astronomy, University  of Cambridge, 
Madingley Road, CB3  0HA, Cambridge, UK. \email{mbergema@ast.cam.ac.uk}
\and
T. Nordlander \at Division of Astronomy and Space Physics, Department of Physics
and Astronomy, Uppsala University, 
Box 516, 75120 Uppsala, Sweden. \email{thomas.nordlander@physics.uu.se}}
%
%
%
\maketitle

\abstract*{xxx}
\abstract{The interpretation of observed spectra of stars in terms of
fundamental stellar properties is a key problem in astrophysics. For FGK-type
stars, the radiative transfer models are often computed using the assumption of
local thermodynamic equilibrium (LTE). Its validity is often questionnable and
needs to be supported by detailed studies, which build upon the consistent
framework of non-LTE. In this review, we outline the theory
of non-LTE. The processes causing departures from LTE are introduced
qualitatively by their physical interpretation, as well as quantitatively by
their impact on the models of stellar spectra and element abundances. We also
compile and analyse the most recent results from the literature. In particular,
we examine the non-LTE effects for 24 chemical elements for six late-studied
FGK-type stars.}

\section{Introduction}

Local thermodynamic equilibrium (LTE) is a common assumption when
solving the radiative transfer problem in stellar atmospheres. The reason for
adopting LTE is that it tremendously simplifies the calculation of number
densities of atoms and molecules. However, the trouble is that the assumption
essentially implies that stars do not radiate. Quoting
\citet{1973ARA&A..11..187M}, 'Departures from LTE occur simply because stars
have a boundary through which photons escape into space'. In what follows, we
will take a short excursus into why this happens and provide a brief overview of
the recent developments in the field. 

We start with reviewing the theoretical foundations of radiative transfer as it
happens in
reality, i.e. including its impact on the properties of matter in stellar
atmospheres. This approach is known as non-local thermodynamic equilibrium
(non-LTE). We discuss the non-LTE 'mechanics' and summarise the impact on
spectral line formation. Finally, we present a list of non-LTE abundance
corrections for the chemical elements, for which detailed statistical
equilibrium calculations are available in the literature.

\section{The framework and notation}
Let us for simplicity take a one-dimensional hydrostatic model of a
stellar atmosphere, which is some relation of temperature and gas
pressure with depth, and consider the transport of radiation outward.

At each depth point, we then assume that matter particles (ions, atoms,
electrons, molecules) are in \emph{local thermodynamic equilibrium} with each
other. This equilibrium is established by intra-particle collisions, resulting
in the Maxwellian velocity distribution and the Saha-Boltzmann
distribution of particles over degrees of excitation and ionisation.
Thus the fundamental assumption of LTE is that particle collisions establish the
energy distribution of matter. The energy distribution of the radiation field
may depart from LTE, $J_\nu \ne B_\nu$, but the influence of non-blackbody
radiation field\footnote{A non-blackbody spectral energy distribution is in
itself non-LTE.} on the energy partitioning of matter is ignored. The local rate
of energy generation $\eta_\nu$ depends only on the local thermodynamic
properties of the gas, i.e. the local kinetic temperature $T_e$ and local
pressure. Therefore the local source function is simply the Planck function
$B_\nu$.

However, the physical truth is that photons and matter particles
interact in a variety of ways: through photo-excitation and ionisation,
radiative recombination, stimulated emission, and other processes. Deep inside
the star, typically far below the photosphere, the collision rates are very
intense and the photon mean free path $\mfp$ is smaller than the scale over
which the physical variables (temperature, pressure) change, so that
the radiation field thermalises to the Planck function, $J_\nu \approx B_\nu$.
Closer to the stellar surface, $\mfp$ becomes large, larger than the scale
height of material. Thus, as photons diffuse outward, their decoupling from
matter increases: the radiation field becomes non-local, anisotropic, and
strongly non-Planckian. Numerical calculations show that the radiative
transition rates driven by the non-local radiation field in the outer
atmospheric layers far outweigh collisional transition rates, thus the
populations of atomic energy levels differ from the LTE values.

\begin{figure}[t]
\begin{center}
\includegraphics[scale=.4]{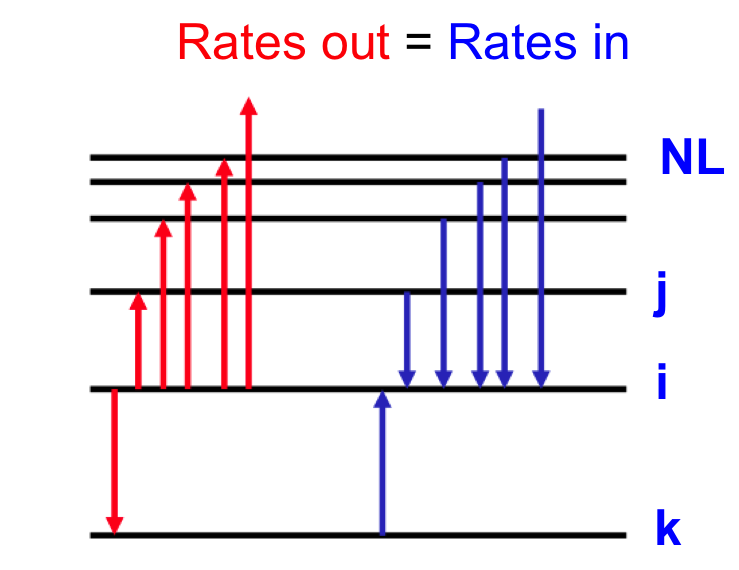}
\caption{Illustration of the statistical equilibrium equation for a level $i$ in
the
model atom; the sum of the rates of all transitions into a level $i$ is balanced
by
the sum of the rates out.}
 \label{fig:f1}
\end{center}
\end{figure}

The non-locality arises because of scattering: photons moving through
the gas only change their direction in a random way and experience a slight
shift in frequency. To account for scattering in the processes of radiative
transfer and, correspondingly, establishment of excitation-ionisation
equilibrium of an element, the concept of non-LTE was introduced.
First, the Saha-Boltzmann equations are replaced by the rate equations or their
time-independent counterpart, statistical equilibrium (SE) or \textit{kinetic
equilibrium} equations. For example, an SE equation for a level $i$, illustrated
in Fig.~\ref{fig:f1}, can be written like:
%
%
%
\begin{equation} 
\label{eq:stateq} n_{i}\sum_{j\neq i}P_{ij} = \sum_{j\neq i}n_{j}P_{ji}.
\end{equation}
Here, $P_{ij} = C_{ij} + R_{ij}$ is the total of collisional and radiative rates
(per particle) which establish the equilibrium number of ions excited to the
level $i$. A subscript $ij$ means that a transition from the level $i$ to the
level $j$ occurs. This equation must be written for each level $i$ of every ion
$c$ in every unit volume of a stellar photosphere, and it describes the
microscopic
interaction between atoms, electrons and photons. 

To close the system of statistical equilibrium equations, the equation of number
conservation can be used for a fixed number density of hydrogen:
\begin{equation}
\label{eq:numcons} \sum_{i,c}n_{i,c} = \frac{\alpha_{\rm el}}{\alpha_{\rm
H}}\left(\sum_{i}n_{i,{\rm H}} + n_p\right)
\end{equation}
where $\alpha_{\rm el}/\alpha_{\rm H}$ is the fraction of all atoms
and ions of a the element relative to that of hydrogen.

For collisional rates, several theoretical formulae exist, also
quantum-mechanical calculations and few experimental data points in the relevant
 energy regime are available. Collisional rates depend only on the local value
of electron temperature and density. For the Maxwellian distribution,
the collision-induced transitions from the lower to the upper level in a
transition are related to the inverse transitions from the upper to the lower
level by the detailed balance principle, i.e. through the Saha-Boltzmann factor
$\exp(-E_i/kT)$ and the level statistical weights. The rates of transitions due
to inelastic collisions are calculated according to:
\begin{equation}\label{eq:col}
C_{ij} = n_e \int \limits_{\upsilon_0}^{\infty}
\sigma_{ij}(\upsilon)\,\upsilon\,f(\upsilon) d\upsilon
\end{equation}
where $\sigma_{ij}(\upsilon)$ is the electron collision cross-section,
$f(\upsilon)$ is the Maxwellian velocity distribution, $\upsilon_0$ is the
threshold velocity with $m \upsilon_0^2/2 = h \nu_0$.

The radiation field enters equation \ref{eq:stateq} through the radiative rates
$R_{ij}$:
\begin{equation} \label{eq:rad}
R_{ij} = \int \limits_{0}^{\infty} \frac{4\pi}{h\nu}
\sigma_{ij}(\nu) J_{\nu} d\nu \hspace{1.5cm} R_{ji} = \int \limits_{0}^{\infty}
\frac{4\pi}{h\nu} U_{ij} \sigma_{ij}(\nu)
\left(\frac{2h\nu^3}{c^2}+J_{\nu}\right)
d\nu\end{equation} 
where $\sigma_{ij}(\nu)$ is the cross-section of a transition from a
state $i$ to a state $j$, i.e. the monochromatic extinction
coefficient per particle in a state $i$. {For bound-bound transitions,
$\sigma_{ij}(\nu) = \sigma^{\rm line}(\nu)$, and for bound-free transitions,
$\sigma_{ij}(\nu) = \sigma^{\rm cont}(\nu)$. The function $U_{ij}$ is 
expressed as:
\begin{equation} U_{ij}  = \left(\frac{n_{i}}{n_{j}}\right)_{LTE}
\exp\left(-\frac{h\nu}{kT}\right)
\end{equation} 
where $j$ is a bound level or a continuum energy state. The mean intensities
$J_{\nu}$ at all frequencies relevant to the transition $i \rightarrow j$ are
derived from the radiative transfer (RT) equation, which can be written in terms
of the optical depth $\tau$:
\begin{equation} \label{eq:optdepth} d\tau_\nu(z) = - \kappa_\nu dz
\end{equation}
\begin{equation} \label{eq:radtran_od} \mu \frac{dI_\nu(\tau_\nu, 
\mu)}{d\tau_\nu} = I_\nu(\tau_\nu, \mu) - S_\nu(\tau_\nu) \end{equation}
For the true (thermal) absorption and emission processes, the source
function S$_\nu$ is defined by the Planck function. In non-LTE, a part of the
local energy emission at a given point is due to the scattering of photons from
the surrounding medium. So the source function also depends on the radiation
field (see Eq. \ref{eq:nltelin_sf_full}), which is at some frequencies
non-local. As a result, we have a coupled system of statistical equilibrium and
radiative transfer equations which must be solved simultaneously to
give the distribution of particles among excitation levels and ionisation
stages. 

The solution of the coupled SE and RT equations is non-trivial and beyond the
scope of this lecture. The most common method is accelerated lambda iteration
\citep{1992A&A...262..209R}.
\section{Non-LTE effects}

\subsection{Statistical equilibrium}

To quantify the departures from LTE, it is necessary to explore relationships of
the kind $\nn/\nl $, i.e. the number density of a given energy level in an atom
or a molecule as computed in non-LTE compared to LTE. This ratio is termed a
departure coefficient, $b_i$. Although there are several definitions, the one
given above \citep{1972SoPh...23..265W} is the most common. Under LTE, $\bi = 1$
and the level is said to be 'thermalised', i.e. its occupation number is just as
that given by the Saha-Boltzmann statistics. Under non-LTE, if $\bi < 1$, the
level is underpopulated, and if $\bi > 1$, the level is overpopulated
relative to its value in LTE.

How do these numbers come about? The algorithm is contained in the statistical
equilibrium equations, which determine the relative populations of the
energy levels to satisfy the given stellar parameters, the gradients of physical
variables in the atmosphere\footnote{Assume that the non-LTE problem is solved
for a fixed model atmosphere structure}, and the intrinsic atomic
properties of an atom or molecule under consideration, i.e the structure of
electronic configurations and the transition probabilities. To understand this
better, we can expand the rates in the SE equation for a level $i$ (Mihalas
1978):
\begin{equation}
\label{eq:expan_stateq} 
\begin{split}
\mathrm{all~transitions~to~a~level~i: ~~~~~} \sum_{j~>~i} n_j \left(
\frac{n_i}{n_j}\right)_{\rm LTE} (R_{ji} + C_{ji} ) + \sum_{k~<~i} n_k (R_{ki} +
C_{ki})  = \\
\mathrm{all~transitions~from~a~level~i: ~~~} n_i \sum_{k~<~i} \left(
\frac{n_i}{n_j}\right)_{\rm LTE} (R_{ik} + C_{ik} ) + n_i  \sum_{j~>~i} (R_{ij}
+ C_{ij}) 
\end{split}
\end{equation}
where $j$ may refer to a bound or a free (ion ground-state or continuum state,
and the radiative and collisional rates depend on the mean intensity (eq.
\ref{eq:rad}). The equations contain both the bound-bound and the bound-free
terms, so the mean intensity must be integrated either over the line profile or
over a large range of energies from the ionisation edge of a given
level.

Neglecting all ionisation processes, equation \ref{eq:expan_stateq}
would simplify to:
\begin{equation}
\label{eq:expan_stateq2} 
\begin{split}
\sum_{j~>~i} n_j \left( \frac{n_i}{n_j}\right)_{\rm LTE} (A_{ji} + B_{ji}
\bar{J} + C_{ji} ) + \sum_{k~<~i} n_k (B_{ki} \bar{J} + C_{ki})  = \\
n_i \sum_{k~<~i} \left( \frac{n_i}{n_j}\right)_{\rm LTE} (A_{ik} + B_{ik}
\bar{J} + C_{ik} ) + n_i \sum_{j~>~i} (B_{ij} \bar{J} + C_{ij})  
\end{split}
\end{equation}
where $\bar{J} = \int \psi_\nu J_\nu d\nu$, $A_{ji}$ and $B_{ji}$ the
Einstein coefficients for spontaneous and stimulated de-excitation (emission),
$C_{ji} $ the rate of collisional de-excitation; the inverse quantities are
represented by the coefficients $A_{ik}$, $B_{ik}$ and $C_{ik}$.

Having solved the equations for all levels in a model atom simultaneously with
the RT equation at all relevant  frequencies, we end up with the
\emph{statistical} (or \emph{excitation-ionisation}) \emph{balance}, which gives
us the non-LTE values of the atomic number densities $n_i$\footnote{In the
literature, the atomic number densities may also be referred to as population
numbers}. The next step is usually the analysis of the individual rates with the
aim to understand how exactly the distribution arises, and in particular to
identify the leading channel which drives the  departures from LTE under the
influence of the radiation field. 

Often, there is no need to analyse each individual rate, as most of them are
very small, so in practice the statistical equilibrium is established by one or
two dominant channels. These channels are, in essence, sequences of processes,
which occur under a certain combination of factors, including the
optical depth in a line or continuum, the difference between $J_\nu$ and the
local Planck function, the relative size of cross-sections for different atomic
levels.

These channels are schematically illustrated in Fig. \ref{fig2}, and they
include:
\begin{figure}[b]
\begin{center}
\includegraphics[scale=0.5]{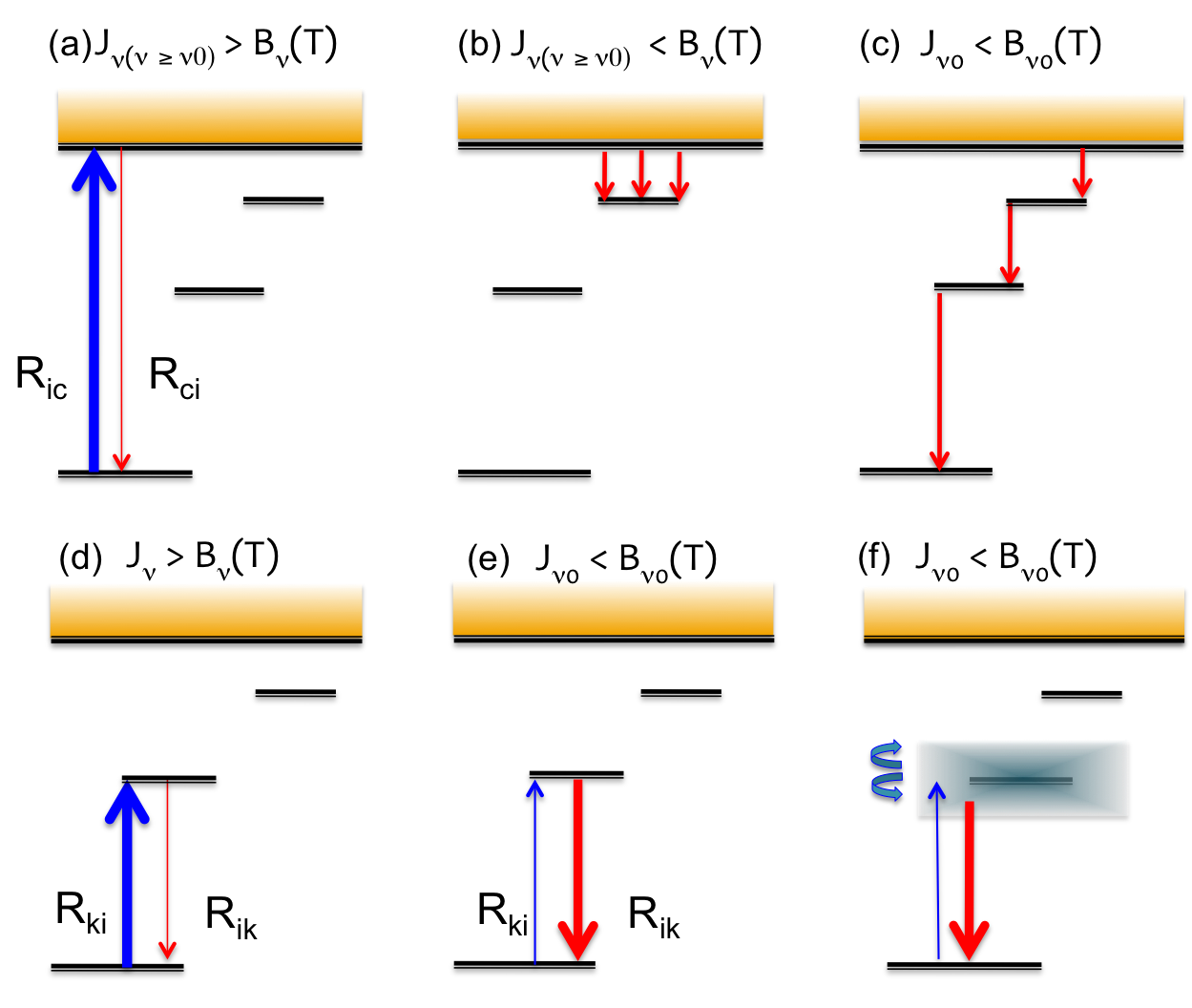}
\caption{Illustration of reaction channels for a hypothetical multi-level atom.
$\nu_0$ is the central frequency for a spectral line or the frequency of
ionisation threshold. See Sec. 3.1}
\label{fig2}
\end{center}
\end{figure}
\begin{figure}[!ht]
\begin{center}
\includegraphics[scale=0.35]{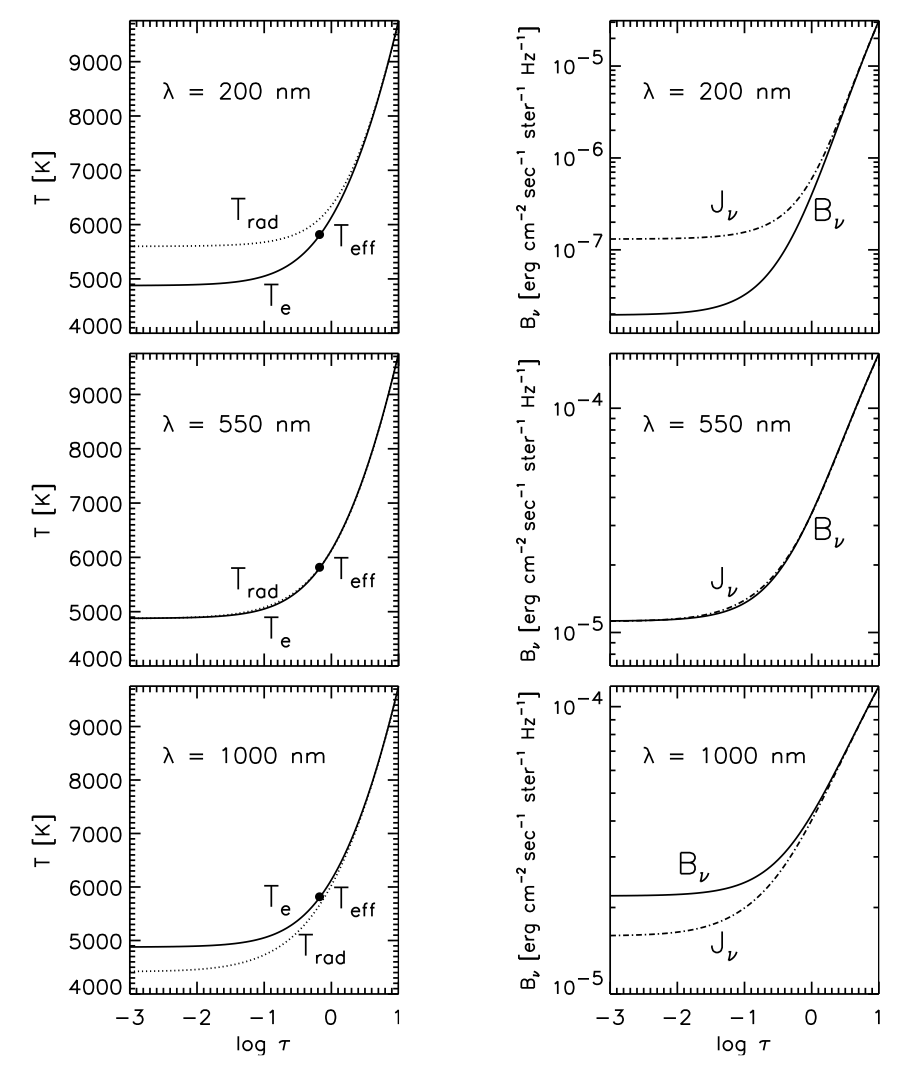}
\caption{Left panel: The comparison of radiation temperatures for different
wavelengths with the local kinetic temperature in a stellar atmosphere. Right
panel: The decoupling of the the mean intensity from the Planck function for
different wavelengths, as indicated in each sub-plot. Figure courtesy Rob
Rutten; \url{http://www.staff.science.uu.nl/~rutte101}.}
\label{fig3}
\end{center}
\end{figure}

\begin{itemize}

\item \textbf{Photoionisation and photon pumping}

A super-thermal radiation field with mean intensity larger than the
Planck function, $J_\nu > B_\nu(T_e)$, leads to over-ionisation (Fig. \ref{fig2}
(a)) and over-excitation (Fig. \ref{fig2} (d)). The latter process is termed
photon pumping. The typical situation is that in a stellar
atmosphere the mean intensity in the continuum $J_{\nu,c}$ lies significantly
above the local Planck function bluewards of the stellar flux maximum.
Thus, over-excitation and over-ionisation affect the neutral atoms (e.g.
\ion{Mg}{I}, \ion{Si}{I}, \ion{Fe}{I}), which have ionisation threshold in the
blue and UV.
\item \textbf{Over-recombination}

If the mean intensity falls below the Planck function, $J_\nu < B_\nu(T_e)$,
the shortage of ionisations leads to the net over-recombination to the upper
levels (Fig. \ref{fig2} (b)), as is the case in the infrared (Fig.
\ref{fig3}, bottom panel). Over-recombination takes place in the bound-free
transitions with the corresponding wavelengths; this is the case for
the majority of neutral atoms (\ion{Na}{I}, \ion{K}{I}, \ion{Cr}{I},
\ion{Fe}{I}, etc).
\item \textbf{Photon 'suction'}

For many elements, the diagram of electronic configurations is very regular,
with a sequence of high-probability transitions connecting the lowest and the
uppermost (Rydberg) levels. This 'ladder' favours photon suction (Fig.
\ref{fig2} (c)), which represents a successive de-excitation  through
high-probability transitions to the lower levels. Photon suction is a sequence
of spontaneous transitions caused by photon loss, in the atmospheric layers
where the line centre optical depth has dropped below unity for every transition
involved in the sequence. Such a sequence can include transitions of various
frequencies, from UV to IR.  Photon suction is balanced by the
recombination flow through the bound-free edges in the IR.
\item \textbf{Photon loss and resonance line scattering}

If the optical depth falls below unity in a line, $\tau_{\nu0} < 1$, the photons
escape without
reconversion to thermal energy. This is termed photon loss. In the
framework of statistical equilibrium, radiative excitations are in deficit, and
spontaneous de-excitations lead to the over-population of the lower level in a
transition out of the upper level  (Fig. \ref{fig2} (e), (f)).  The effect is
particularly important for strong resonance lines (e.g. Na D or Mg b) with very
small collisional destruction probability: photon escape in the line itself
influences the line source function, causing sub-thermal S$_\nu$.
\end{itemize}

Of course, in most cases all these processes form a sophisticated bundle, which
is very difficult to unfold. The  non-LTE effects depend upon several factors,
including the physical conditions and the atomic structure (ionisation energy,
characteristics of the energy levels in the atom, the number of allowed and
forbidden transitions, size of cross-sections). However, generally
atoms fall into one of the two categories: \emph{photo-ionisation dominated}
atoms and
\emph{collision-dominated} ions \citep[see][and references
therein]{2001A&A...366..981G}:

\begin{itemize}
\item \textbf{Photo-ionisation-dominated ions}

The main non-LTE driver is over-ionisation. The typical situation is that
low-lying energy levels of over-ionising species have very large
photo-ionisation cross-sections, and that their ionisation edges are located in
the UV, where $J_\nu > B_\nu(T_e)$. Moreover, the photo-ionisation
cross-sections of atoms with complex electronic configurations often have
prominent resonances at certain energies which further enhances the radiative
rates. For such atoms (e.g., \ion{Mg}{I}, \ion{Al}{I}, \ion{Si}{I}, \ion{Ca}{I},
\ion{Mn}{I}, \ion{Fe}{I}, \ion{Co}{I}), the non-LTE departure coefficients are
generally below unity, i.e. there is a sub-set of strongly under-populated
levels, and this under-population is redistributed over all other levels by
collisions or optically thick line transitions. The non-LTE effects usually grow
with increasing $\teff$, and decreasing $\feh$ and $\log g$, due to stronger
non-local radiation fields in the UV, lesser efficiency of thermalising
collisions, and vanishing line blocking at low metallicity.

\item \textbf{Collision-dominated ions}

With very small photo-ionisation cross-sections, the statistical equilibrium of
collision-dominated ions reflects the interplay between the strong line
scattering, over-recombination, and collisional thermalisation. Representative
ions with these properties are \ion{Na}{I} and \ion{K}{I}. We note, however,
that this is a general picture in 1D hydrostatic model atmospheres, and may not
be valid in the statistical equilibrium calculations with 3D RHD models
\citep[see e.g.][]{2013A&A...554A..96L}.
\end{itemize}

A very detailed discussion of these processes with different thought experiments
and toy model atoms, is given in \citet{1992A&A...265..237B}. We also refer the
reader to \citet{1973ARA&A..11..187M}, who provide an excellent physical basis
for the qualitative understanding of the overall picture.
\subsection{Spectral line formation}

Once we know what non-LTE effects dominate the statistical equilibrium of an
atom, the behaviour of spectral line profiles is easy to understand. The
analysis is aided by the diagrams of level departure coefficients for the levels
involved in the transition as a function of some reference optical
depth, e.g. the optical depth at $500$ nm in the continuum.
For the strong lines, it is often helpful to explore the depths of formation of
various parts of the line profiles\footnote{In the literature, one may often
find expressions of the following kind: A spectral line is formed 'at a given
optical depth'. The usual interpretation of this depth is the location where
$\tau_{\lambda,0} = 1$ (at the wavelength of the line core, $\lambda_0$) is
achieved. Of course, a spectral line forms over an interval of depths. For some
strong lines, this range encompasses the whole stellar atmosphere, i.e. several
hundred km.}, for example the layers where in the core or in the wing
$\tau_\nu^l = 1$.
The emergent intensity within a spectral line is given by the radiative
transfer equation:
\begin{equation} \label{eq:int_linfor} I_\nu(\tau_{\nu} = 0, \mu) = 
\int_{0}^{\infty} S_\nu(\tau_{\nu})e^{-\tau_{\nu}/\mu} d\tau_{\nu}/\mu
\end{equation}
where $\mu = \cos \theta$, for the photons travelling in the direction
specified by the angle $\theta$ with respect to the normal. The atomic levels
populations are contained in the source function $S_\nu$ and in the optical
depth $\tau_\nu$ (eq. \ref{eq:optdepth}). 
The line extinction coefficient (related to $\tau_\nu$ via
equation~\ref{eq:optdepth}), or opacity, is defined by:
\begin{equation} \label{eq:nltelin_op_full} 
\kappa_\nu^l \propto n_i\left(1 - \dfrac {n_j g_i}{n_i g_j} \right) = b_i
\left(1 - \dfrac {b_j}{b_i}e^{-\frac{h\nu}{kT}} \right)
\end{equation}
In the Wien's regime ($h\nu > kT$), the expression reduces to: 
\begin{equation} \label{eq:nltelin_op} \kappa_\nu^l \propto b_i 
\end{equation}
When $b_i < 1$, the line opacity is decreased and thus the line formation depth
increases. As temperature is higher in the deeper layers, the result is a
weakening of the absorption line. The other way around, $b_i > 1$ leads to an
increased opacity in the line over an LTE value. The line thus forms in the
outer atmospheric layers, where temperatures are lower, which leads to the line
strengthening with respect to the continuum. 

The line source function can be written as:
\begin{equation} \label{eq:nltelin_sf_full}
S^l = \dfrac{2h\nu^3}{c^2} \dfrac{1} {\dfrac {n_i g_j}{n_j g_i} - 1} =
\dfrac{2h\nu^3}{c^2} \dfrac{1}{\dfrac {b_i}{b_j}\left(e^{\frac{h\nu}{kT}} -
\dfrac{b_j}{b_i}\right)} 
\end{equation}
where we make the assumption of complete redistribution in a line
profile, which implies that the frequency and direction of an absorbed and
emitted photons are not correlated.
When $h\nu > kT$ (in the UV and visual spectral range): 
\begin{equation} \label{eq:nltelin_sf}
\dfrac{S^l}{B_\nu^l} \approx \dfrac{b_j}{b_i}
\end{equation}
A spectral line is weakened when $b_j > b_i$. The situation arises when $S^l >
B_\nu^l$, implying there are more atoms excited to the upper level of a
transition $j$ compared to those on the lower level $i$, than would
result under LTE. Hence, there is more emission and less absorption in the line
in comparison with the Planck's law. Conversely, the absorption line
strengthens when $b_j < b_i$, or $S^l < B_\nu^l$, as there are not enough atoms
on the level $j$ to maintain the same emissivity as in LTE.

Weak lines with small opacity relative to the continuum are formed in the deep
atmospheric layers. Their intensity profiles simply reflect the profile of
$\kappa_\nu$. Thus whatever process, i.e over-ionisation or over-excitation,
causes smaller line opacity, $b_i < 1$, will result in a weaker absorption
line. 

Strong lines influence their own radiation field and also the
cross-talk between different frequencies within the line is important. The
simplest case is that of strong resonance lines, which have a nearly
two-level-atom source function. As described above, the surface photon losses
cause $J_\nu < B_\nu$ and sub-thermal source function $S^l < B_\nu$, that leads
to darker line cores compared to LTE.

In general, there is no simple analytic solution for estimating the effect of
$b_i$ inequalities on line profiles. Non-LTE effects may change the
equivalent width of a line, but its profile may also be affected, even
if the equivalent width is conserved. Thus, departures from LTE may
appear in both curve-of-growth and spectrum synthesis methods, with
sometimes subtle differences.
\section{The impact of non-LTE on elemental abundance analyses}\label{sec:final}
The effects of non-LTE on the abundance determinations in cool stars have been
presented in several excellent review articles, perhaps the most up-to-date and
comprehensive study is that by \citet{2005ARA&A..43..481A}. There is little that
can be added to these papers. We thus restrict the discussion to representative
groups of chemical elements with similar non-LTE effects, and provide a set of
plots, which can be used to gain a qualitative understanding of how the non-LTE
abundance corrections vary with
stellar parameters. These corrections are shown in Fig. \ref{fig5} for six
well-studied stars with parameters listed in Table \ref{Tab1}: the Sun, Procyon,
Arcturus, HD 84937, HD 122563, and HD 140283. The references to the data are
given in Table \ref{table:nlte} in the Appendix. Figure~\ref{fig6} also shows
the results for the individual chemical elements.
\begin{svgraybox}
Non-LTE abundance correction for a given chemical
element,  $\Delta_{\rm El}$, is
defined as, $ \Delta_{\rm El} = \log \rm{A (El)}_{\rm non-LTE} - \log \rm{A
(El)}_{\rm LTE} $, i.e., it is the logarithmic correction, which has to be
applied to an LTE abundance determination A of a specific line to obtain the
correct value corresponding to the use of non-LTE line formation. 
\end{svgraybox}

Several interesting observations can be made based on these plots. For the lines
of the photo-ionisation type ions (\ion{B}{I}, \ion{Mg}{I}, \ion{Al}{I},
\ion{Si}{I}, \ion{Ca}{I}, \ion{Sr}{I}, as well as all neutral Fe-peak
atoms: \ion{Ti}{I}, \ion{Cr}{I}, \ion{Mn}{I}, \ion{Fe}{I}, \ion{Co}{I},
\ion{Ni}{I}) the non-LTE abundance corrections increase with decreasing
metallicity. This tendency simply reflects the fact that the atoms experience a
larger degree of over-ionisation and  radiative pumping due to stronger $J_\nu -
B_\nu$ splits in the low-metallicity atmospheres. The spectral
lines are weaker in non-LTE compared to LTE, and the non-LTE abundances are
correspondingly larger.

For the collision-type minority ions, like \ion{Na}{I} and \ion{K}{I}, the
non-LTE effects on the line formation are mainly due to photon suction
and over-recombination. The spectral lines are typically strengthened, with
non-LTE abundance corrections in the range from $-0.1$ to $-0.5$ dex (Fig.
\ref{fig6}). The efficiency of over-recombination, caused by $J_\nu <  B_\nu$ in
the IR, grows with increasing $\teff$, which is why departures from LTE have the
largest effects for warm turn-off stars, like Procyon.

The diagnostic lines of atoms like \ion{Li}{I} (671 nm), \ion{C}{I} (910, 960
nm), \ion{O}{I} (777 nm triplet), \ion{Ca}{II}, \ion{Sr}{II}, and \ion{Ba}{II},
are often stronger in non-LTE. However, the effect depends on stellar
parameters, elemental abundance, and the atomic data for a transition
\citep[see also][]{1974ApJ...187..147J}. At higher metallicity, or elemental
abundance, the strong lines have sub-thermal source functions due to
photon loss through the wings. The non-LTE abundance corrections are negative
for \ion{C}{I} and \ion{O}{I}, as well as for \ion{Sr}{II} and \ion{Ba}{II} at
solar $\feh$. However, the non-LTE effects for \ion{Sr}{II} and \ion{Ba}{II}
drastically change at  low-metallicity ($\feh < -2$): the lines tend to be
weaker in non-LTE and the abundance corrections grow positive. For 
\ion{Eu}{II}, the positive non-LTE corrections are caused by radiative pumping
which weakens the lines, especially in metal-poor stars.
 
Note that the lines of atoms affected by hyperfine structure (e.g. \ion{Mn}{II},
\ion{Ba}{II}) should be treated with HFS components also in the non-LTE line
formation calculations. This may turn out to be important for higher-metallicity
stars.

\renewcommand{\tabcolsep}{3pt}
\ctable[caption={Stellar parameters for the selected reference stars. $\teff$
and $\logg$ have been independently determined by fundamental relations, while
$\feh$ is the typical literature value.},
label={Tab1}, mincapwidth=0.95\textwidth]{l ccr}{}
{ \FL
Star & $T_{\rm eff}$ & $\log g$ & [Fe/H] \ML
Sun           & $5777$ &  $4.44$ & $0.00$  \NN
Procyon    & $6545$ &  $3.99$ & $-0.05$ \NN
Arcturus    & $4247$ &  $1.59$ & $-0.52$ \NN
HD 84937     & $6275$ & $4.11$  & $-2.00$ \NN
HD 140283    & $5720$ & $3.67$  & $-2.50$ \NN
HD 122563    & $4578$ &  $1.61$  & $-2.74$ \LL
}
%
%
\newcommand\scale{0.55}
\begin{figure}[!ht]
\begin{center}
\includegraphics[scale=\scale]{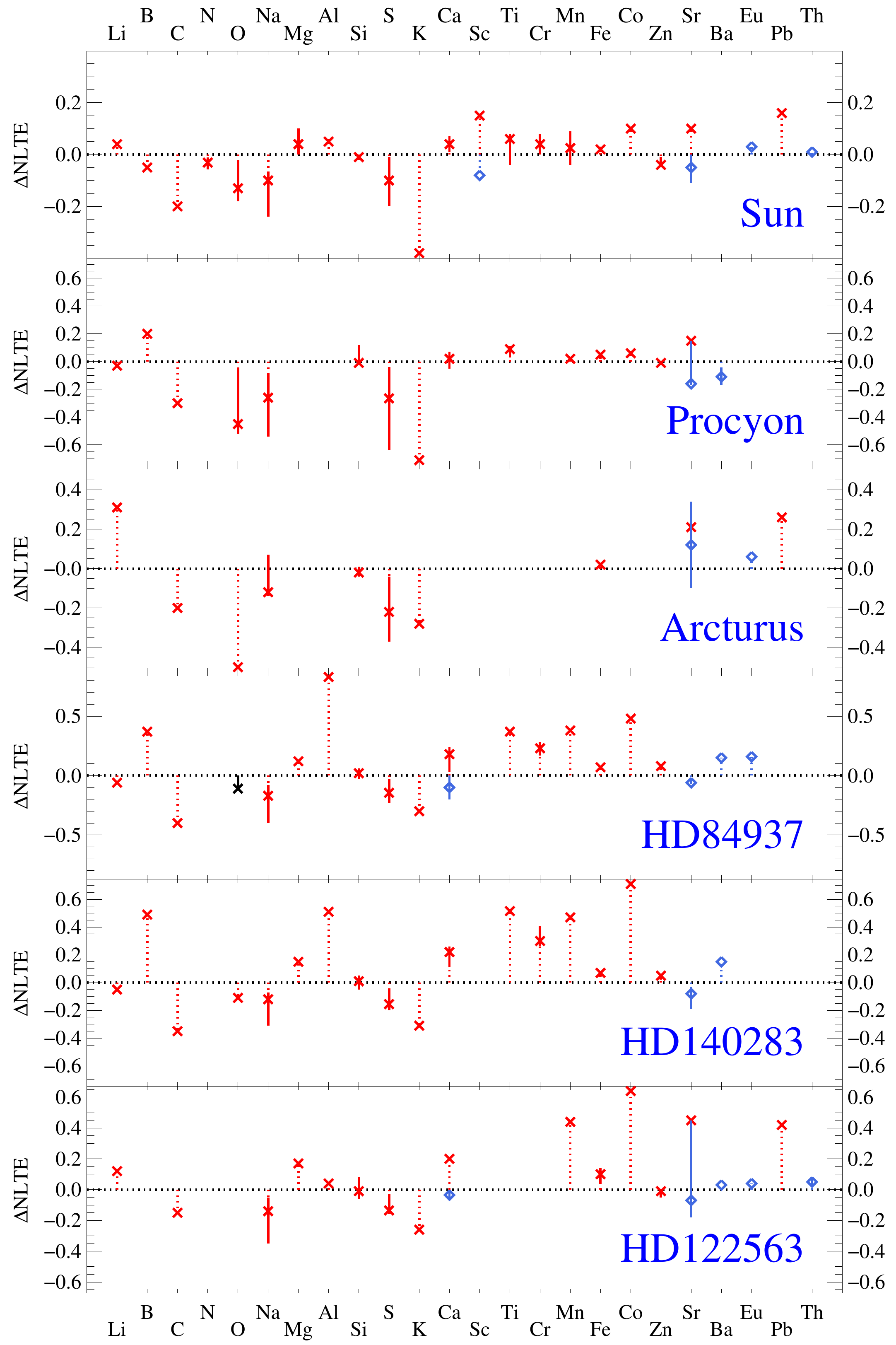}
\caption{Non-LTE effects on 24 elements for six representative, well-studied
stars. Non-LTE (NLTE) effects are given as abundance corrections $\Delta {\rm
NLTE} = A(X)_{\rm NLTE} - A(X)_{\rm LTE}$. Red crosses and blue diamonds show
results obtained from the lines of neutral and ionised atoms, respectively.
Solid, vertical lines indicate min--max ranges of the corrections.
The \ion{O}{I} non-LTE correction for Arcturus is the limiting value taken from
\citet[Fig.~5]{2013AstL...39..126S}. For HD 84937 we adopt the same value as
for HD 140283, since \citet{reetz} found the same values for stars with similar
parameters.}
 \label{fig5}
\end{center}
\end{figure}

\begin{figure}[p]
\begin{center}
\includegraphics[scale=\scale]{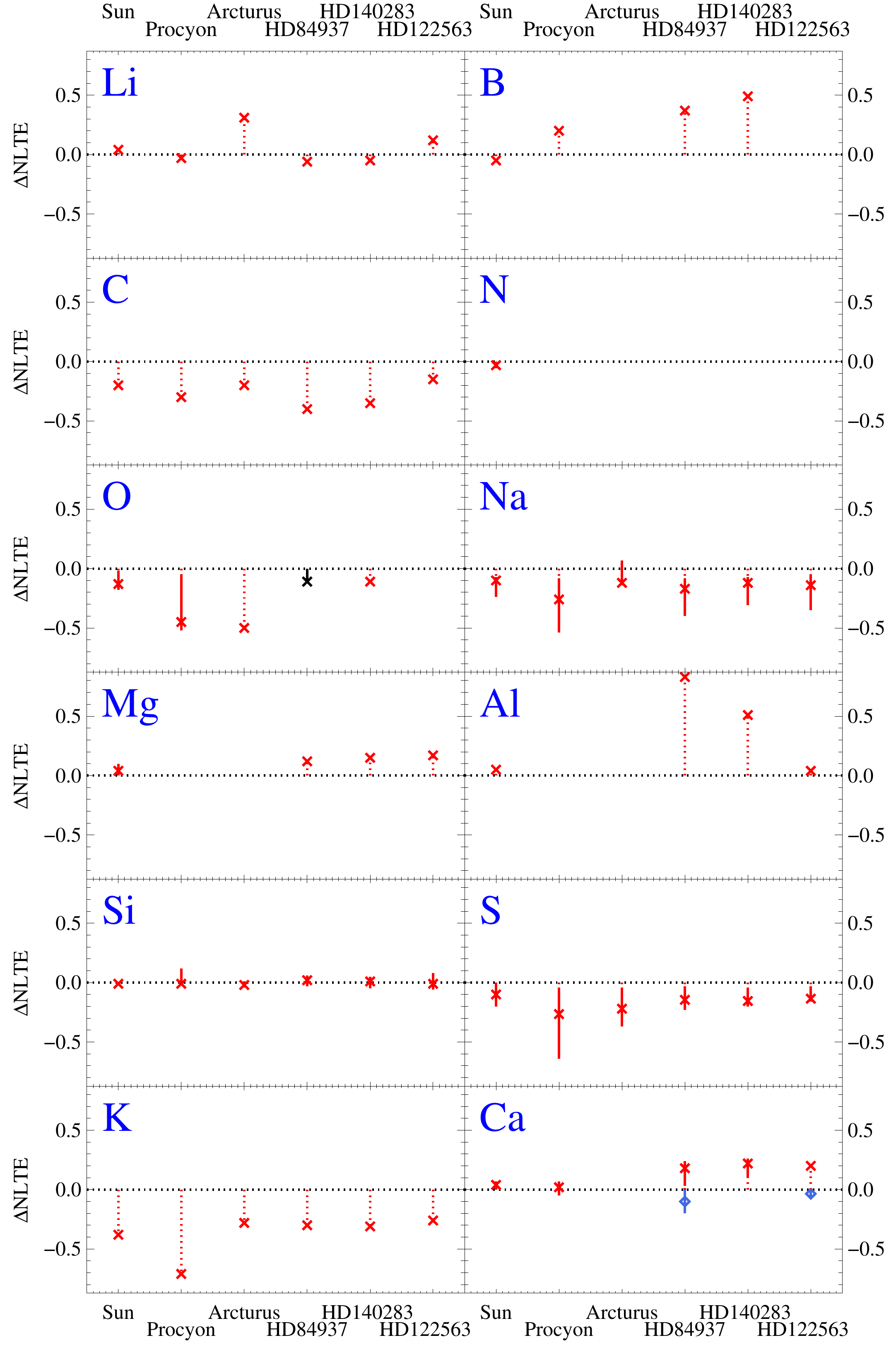}
\end{center}
\caption{As Fig.~\ref{fig5}, but element-by-element.}
\label{fig6}
\end{figure}

\begin{figure}[p] \samenumber
\begin{center}
\includegraphics[scale=\scale]{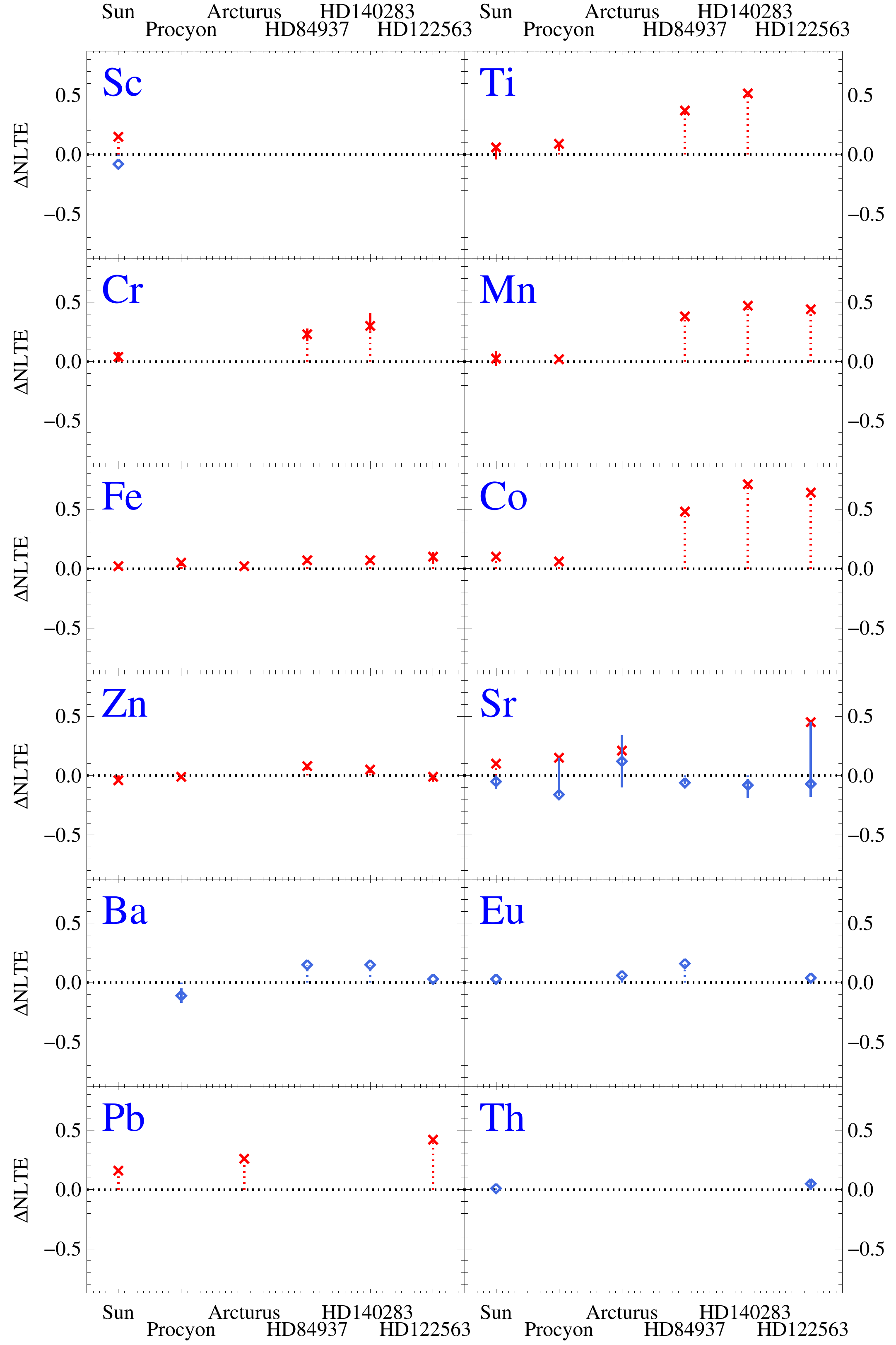}
\end{center}
\caption{\textit{continued.}}
\label{fig7}
\end{figure}
\section{Summary and conclusions}

Non-local thermodynamic equilibrium is a framework, which describes consistently
the
propagation of radiation in a stellar atmosphere and its coupling to matter. LTE
is the boundary case of non-LTE, the approximation in the limit of infinitely
large collision rates. 

In this review lecture, the focus is on non-LTE conditions in the atmospheres of
cool late-type stars. The impact of non-LTE on stellar parameter and abundance
determinations can be gained from detailed theoretical and observational
analyses, and these are now available for the most important chemical elements
observed in the spectra of cool stars. So far, most studies in the literature
focussed on deviations from LTE in the spectral lines of neutral or
singly-ionised atoms. 

Generally, the results obtained by independent groups and methods are consistent
and they can be summarised as follows. There are several well-defined types of
non-LTE effects, and consequently several groups of species which behave
similarly under the same physical conditions in a stellar atmosphere (i.e. given
the same $\teff$, $\log g$, and $\feh$). The first group is formed by species
sensitive to over-ionisation (e.g. \ion{Mg}{I}, \ion{Si}{I}, \ion{Ca}{I},
\ion{Fe}{I}), the second group are collision-dominated species (\ion{Na}{I},
\ion{K}{I}), and the rest are mixed-type ions (e.g. \ion{Li}{I}, \ion{O}{I},
\ion{Ba}{II}), which may show positive or negative non-LTE effects depending
upon different factors. 

In some cases, the non-LTE abundance corrections vary by an order of magnitude
between the groups of elements. In other cases, cancellation may occur such that
even the LTE analysis may
provide the correct abundance \emph{ratios} of two elements. Such pairs
are difficult to establish, although, to first approximation, one may form pairs
from elements in the same group, taking into account the excitation potential of
the line as well as the elemental abundances. We want to stress, however, that a
widely-spread 'rule of thumb' that the lines of the same ionisation stage can be
safely modelled in LTE is a misconception.

Generally, any element analysed under the assumption of LTE should be
regarded with caution (at least) until careful non-LTE studies have been
conducted. Even then, second-order non-LTE effects related to line strengths and
excitation potentials may turn out to fundamentally alter the outcome. With
these potential pitfalls, we urge the reader to take non-LTE corrections into
account whenever possible.

\begin{acknowledgement}
Fig. 3 has been kindly provided by Rob Rutten.
\end{acknowledgement}

\bibliographystyle{spbasic}
\bibliography{references,nlterefs}
\appendix

\ctable[label={table:nlte}, caption={References to NLTE studies of late-type
stars in the literature, which were used to  create Figs.~\ref{fig5} and
\ref{fig6}.}, maxwidth=\textwidth]
{ll <{\raggedright}X <{\raggedright}X} 
{
\tnote[a]{\ion{H}{I} collision rates for ionisation from
\citet{1991JChPh..95.5738K}; no b-b collisions included in the model atom.}
\tnote[b]{Based on quantum mechanical calculations of hydrogen collision rates}
\tnote[c]{Results were extracted using \url{www.inspect-stars.net}}
\tnote[d]{$S_H$ set as exponentially dependent on upper excitation energy}
\tnote[e]{But $S_H = 0.002$ for levels where $n > 7$, affecting IR lines.}
\tnote[f]{Corrections were computed specifically for this work}
}
{ \FL
Ion & $S_H$ & Reference & Spectral lines [nm] \ML
\multicolumn{4}{c}{Light elements} \ML
\ion{B}{I} & x\tmark[a] & \citet{1996AA...311..680K} 
    & 250 \NN
\ion{Li}{I} & QM\tmark[b] & \citet{2009AA...503..541L}\tmark[c] 
	& 670.7 \NN
\ion{C}{I} & 0 & \citet{2006AA...458..899F} 
    & 909.5 \NN
\ion{N}{I} & $1/3$ & \citet{2009AA...498..877C} 
    & 12 lines, \wlrange{744}{1054} \NN
\ion{O}{I} & 0 & \citet{reetz} 
    & 777.1--777.5 \NN
\ion{O}{I} & 1 & \citet{2013AstL...39..126S} 
    & 615.8, 777.1--777.5, 844.7 \ML
\multicolumn{4}{c}{Intermediate elements} \ML
\ion{Na}{I} & QM\tmark[b] & \citet{2011AA...528A.103L}\tmark[c] 
	& 568.8, 589.5, 615.4, 819.4\NN
\ion{Mg}I & x\tmark[d] & {\citet{1998A&A...333..219Z}}
    & 13 lines, \wlrange{457}{892} \NN
\ion{Mg}{I} & 0.05 & \citet{2006AA...451.1065G} 
	& 7 lines, \wlrange{457}{571} \NN 
\ion{Mg}I & 0.1 & \citet{2008AA...478..529M}
	& 6 lines, \wlrange{457}{571} \NN 
\ion{Al}I & 0.4\tmark[e] & {\citet{1997A&A...325.1088B}}
    & 396.2, 669.7 \NN
\ion{Al}{I} & 0.002 & \citet{2006AA...451.1065G,2008AA...478..529M} 
	& 396.2 \NN
\ion{Si}{I} & 1 & \citet{2013ApJ...764..115B}\tmark[f]; \citet{wuerl} 
	& 9--10 lines, \wlrange{390}{723} \NN
\ion{S}{I} & 1 & \citet{2005PASJ...57..751T} 
	& 8 lines, \wlrange{869}{1046} \NN 
\ion{K}{I} &  0.001 & \citet{2002PASJ...54..275T} 
	& 769.9\NN
\ion{Ca}I & 0.1 & \citet{2007AA...461..261M,2008AA...478..529M} 
    & 422.6, 442.5, 526.1, 534.9, 616.2 \NN
\ion{Sc}{I} & 0.1 & \citet{2008AA...481..489Z}     & 567.2 \NN \ion{Sc}{II} &
0.1 & \citet{2008AA...481..489Z}     & 552.6 \NN
\ion{Ti}{I} & 0.05 & \citet{2011MNRAS.413.2184B} 
    & 453.4, 498.1, 502.2, 843.5  
\ML
\multicolumn{4}{c}{Fe-group elements} \ML
\ion{Cr}{I} & 0 &  \citet{2010AA...522A...9B} 
    & 425.4, 520.6, 540.9 \NN
\ion{Mn}{I } & 0.05 & \citet{2008AA...492..823B} 
    & 475.4, 467.1, 601.6, 874.0 \NN 
\ion{Fe}{I } & 1 & \citet{2012MNRAS.427...27B, 2012MNRAS.427...50L}\tmark[c] 
    & 492.0, 499.4, 521.6, 524.2, 606.5, 625.2, 643.0 \NN
\ion{Co}{I } & 0.05 &  \citet{2010MNRAS.401.1334B} 
    & 350.1, 411.0, 412.1 \NN 
\ion{Zn}{I}  & 1 &   \citet{2005PASJ...57..751T} 
	& 472.2, 481.1, 636.2\ML
\multicolumn{4}{c}{Neutron-capture elements} \ML
\ion{Sr}{I} & 0 & \citet{2012AA...546A..90B}\tmark[c]     & 460.7 \NN
\ion{Sr}{II} & 0 & \citet{2012AA...546A..90B}\tmark[c]     & 407.7, 421.5, 1004,
1033, 1091  \NN
\ion{Pb}{I} & 0.1 & \citet{2012AA...540A..98M} 
    & 405.7 \NN
\ion{Th}{II} & 0.1 & \citet{2012AA...540A..98M} 
    & 401.9, 408.6 
\NN
\ion{Eu}{II} & 0 & \citet{2008AA...478..529M,2012AA...540A..98M} 
    & 413.0 \NN
\ion{Ba}{II} & 0 & \citet{1999AA...343..519M,2008AA...478..529M} 
    & 455.4, 585.3, 649.6 
1\&3) 
\LL
}

\end{document}